# A fiber-bundle model for the continuum deformation of brittle material

K. Z. Nanjo[1,2]

[1]Address: Global Center for Asian and Regional Research, University of Shizuoka

3-6-1, Takajo, Aoi, Shizuoka, 420-0839, Japan

Tel: +81-54-245-5600

Fax: +81-54-245-5603

E-mail: nanjo@u-shizuoka-ken.ac.jp

[2]Address: Institute of Advanced Sciences, Yokohama National University

79-5, Tokiwadai, Hodogaya, Yokohama 240-8501, Japan

**Abstract** The deformation of brittle material is primarily accompanied by micro-cracking and faulting. However, it has often been found that continuum fluid models, usually based on a non-Newtonian viscosity, are applicable. To explain this rheology, we use a fiber-bundle model, which is a model of damage mechanics. In our analyses, yield stress was introduced. Above this stress, we hypothesize that the fibers begin to fail and a failed fiber is replaced by a new fiber. This replacement is analogous to a micro-crack or an earthquake and its iteration is analogous to stick-slip motion. Below the yield stress, we assume that no fiber failure occurs, and the material behaves elastically. We show that deformation above yield stress under a constant strain rate for a sufficient amount of time can be modeled as an equation similar to that used for non-Newtonian viscous flow. We expand our rheological model to treat viscoelasticity and consider a stress relaxation problem. The solution can be used to understand aftershock temporal decay following an earthquake. Our results provide justification for the use of a non-Newtonian viscous flow to model the continuum deformation of brittle materials.

# 1 Introduction

Material fracture is a complicated phenomenon. Even if the material appears to be homogenous, there will be a distribution of dislocations, flaws, and other heterogeneities present. As the applied stress is increased, uncorrelated micro-cracks occur randomly on the heterogeneities. As the density of micro-cracks increases, the stress fields of the micro-cracks interact and the micro-cracks become correlated. The micro-cracks eventually coalesce to form a through-going fracture. Even in an idealized case where propagation of a single fracture goes through a homogenous solid, this is poorly understood by dynamic fracture mechanics because of the singularities at the crack tip (Freund 1990). However, this irreversible process can be treated as a part of damage mechanics. Generally, the irreversible deformation of a solid is referred as "damage" (Kachanov 1986; Krajcinovic 1996). Thus, all deformation associated with decohesion between inclusions, accumulation of dislocations leading to the nucleation of micro-cracks, debonding of fibers and matrix in composite materials and other events can be covered by this term.

Brittle and ductile deformation plays important roles in the irreversible deformation of a solid. In the brittle process, significant deformation is localized at planar surfaces, on which a through-going fracture is formed, as described above. Other examples of brittle deformation are associated with displacements on faults in the Earth's crust (e.g. King 1983; King et al. 1994; Thatcher 1995; Jackson 2002) and stick-slip motion between elastic bodies (e.g. Yamaguchi et al. 2009; Morishita et al. 2010) and other materials.

An alternative approach to irreversible deformation is the ductile deformation, which can be described by utilizing creep (Newtonian and non-Newtonian) and plastic rheology. An empirical power-law equation between strain rate $\dot{\varepsilon}$ and stress $\bar{\sigma}$, often calle Dorn's equation, has been widely used (e.g. Dorn 1954; Nicolas and Poirier 1976; Karato and Wu 1993)

$$\dot{\varepsilon} = A\bar{\sigma}^n exp\left(-\frac{Q+PV_a}{RT}\right), \qquad (1)$$

where $A$ and $n$ are constants, $Q$ is the activation energy, $V_a$ is the activation volume, $R$ is the gas constant, $T$ is the absolute temperature, and $P$ is the pressure. It is valid for both diffusion creep ($n$ = 1) and dislocation creep ($n$ = 3-5) which are thermally activated.

Both brittle and ductile processes can be quantified using the concept of dislocations. Dislocation theories have been developed by many scientists to explain not only the mechanical properties but also the optical and electromagnetic properties of crystals. Taylor (1934), Orowan (1934) and Polanyi (1934) applied dislocation studies to the plastic deformation of simple crystals whose dislocation is much lower than the theoretical values calculated from atomic theory assuming a perfect-lattice state. Mura (1969) developed a method of continuously distributed dislocations to consider the relationship between macroscopic plasticity and dislocation theory. However, these studies have not considered how micro-cracks contribute to the deformation of solids. Deformation associated with micro-cracks, stick-slips, and faults can be treated as dislocations, which are shear deformations across planar surfaces.

Linking between large-scale or long-term ductile deformation to small-scale or short-term brittle process is of great interest in engineering, physics, material science, and geophysics (e.g. Sornette and Virieux 1992; Ma and Kuang 1995; Kovács et al. 2013; Alava et al. 2006; Lyakohovsky et al. 1997; Kun et al., 2006; Hansen et al. 2015). For example, Sornette and Virieux (1992) theoretically derived a link between short-timescale deformation due to slips on faults to long-timescale tectonics. Computer simulation of composite materials (Kovács et al. 2013) showed how brittle failure at the microscopic level leads to a ductile macroscopic response. An avenue for irreversible behavior associated with brittle and ductile processes is damage mechanics. The concept of damage mechanics has been utilized in resolving engineering problems (Krajcinovic 1996; Skrzypek and Ganczarski 1999; Voyiadjis and Kattan 1999). Two models that have been utilized to do this are the fiber-bundle model (FBM) and the continuum damage model (CDM).

CDM is widely used in civil and mechanical engineering (e.g. Kachanov 1986) and also applied to Earth's tectonic processes (e.g. Lyakhovsky et al. 1997; Ben-Zion and Lyakhovsky 2002). A damage variable $\alpha$ is introduced to quantify deviation from linear elasticity and the distribution of micro-cracks in the material being considered. By definition, $\alpha$ assumes values between 0 and 1 and failure occurs when α = 1. Damage evolution is a transient process so that we have α(*t*) until failure occurs (α = 1).

Another approach to the irreversible deformation of materials is provided by FBM, a discrete model of damage mechanics. FBM was applied to fatigue in structural materials and earthquakes in geophysical settings (e.g. Hemmer and Hansen 1992; Moreno et al. 2001; Hansen et al. 2015). The behavior of irreversible deformation is quantified by the original number of fibers in the bundle $n_0$ and the number of remaining fibers $n_f$. Damage evolution is a transient process of $(n_0 - n_f)/n_0$ from 0 to 1 (as an equivalent, the decrease of $n_f$ from $n_0$ to 0). CDM was shown to be equivalent to FBM for assessing the occurrence of a failure in a simple geometry (Krajcinovic 1996; Turcotte et al. 2003; Turcotte and Glasscoe 2004).

Our aim is to explain the continuous deformation of heterogeneous solid material with dislocations. Continuously deformed material is hypothesized to include heterogeneity that influences motion during dislocations. We assume that displacements on dislocations dominate. We show that when FBM is applied to the brittle deformation of a solid, a non-Newtonian power law viscous rheology is obtained. In our analyses, yield stress is introduced as follows: below this stress the solid behaves elastically and can act as a stress guide, and above this stress the continuum deformations can be modeled as a power-law viscous fluid. Yield stress needs to be defined in a transition from brittle or elastic behavior to ductile or plastic behavior. This expands upon previous studies (Turcotte and Glasscoe 2004; Nanjo and Turcotte 2005) that did not consider yield stress.

As an application example, we use a viscoelastic version of our model under a constant strain applied to a sample. We show that because of damage, the stress on the sample relaxes and this relaxation can reproduce the power-law temporal decay. Our result shows good agreement with Omori-Utsu's law for aftershock decay following an earthquake (Omori 1894; Utsu 1961).

Our rheological model, which we will show in this paper, is generic to understand the continuum deformation of brittle materials in not only geophysics but also engineering and material science. We believe that the solution to a relatively simple problem in this paper is illustrative.

**2 FBM**

2.1 Fiber Failure Criterion

While FBM strives to establish a link between microscopic deformation features and macroscopic observations, the model is phenomenological on many levels (e.g. in the distribution rules). Thus, the first requirement is to specify the failure criterion for the fibers. When stress is applied to a fiber bundle, the fibers begin to fail. Local failure events are usually dynamic in reality so that inertial effects should be taken into consideration during episodic fiber failures. Therefore, we used the dynamic time-dependent failure model where the distribution of failure time of fibers is specified in terms of the stress on the fibers (Coleman 1958). The alternative model is static in that the distribution of the failure strengths of the fiber is specified (Daniel 1945).

It is also necessary to specify how the stress on a failed fiber is redistributed to the remaining fibers (Smith and Phoenix 1981; Kun et al. 2006). Continuum mechanics established rules for interaction between forces and deformations at distances (e.g. long range elastic interaction, force field due to a dislocation, dipoles, quadruples, and so on). In order to treat this long range interaction, we used the uniform load-sharing hypothesis in which the stress from a failed fiber is redistributed equally to the remaining fibers (e.g. Hemmer and Hansen 1992; Turcotte et al. 2003). The alternative is the local load-sharing hypothesis, where stress from a failed fiber is redistributed to neighboring fibers (usually nearest neighbors) (Newman and Phoenix 2001). We understand that the local load sharing and the uniform load sharing are two extreme forms of the load sharing rule. In continuum mechanics of elastic materials, the stress distribution around cracks follows a power law relation between stress increase and distance from the crack tip (e.g. Lawn and Wilshaw 1975). Motivated by this result of fracture mechanics (Kun et al. 2006), an important future work is to extend FBM by introducing a load-sharing rule of the power-law form, which we do not address in this paper.

In order to apply FBM to a continuously deforming solid material, a failed fiber is replaced by a new fiber. The fiber replacement hypothesis has been previously used by Zapperi et al. (1997), Kun et al. (2000), Moreno et al. (2001), and Halász and Kun (2009). We hypothesize that the replacement of a broken fiber by a new fiber is analogous to an earthquake rupture, a micro-crack or the migration of a dislocation. Replacing fibers allows us to model the repetitive occurrence of earthquakes on a fault, migration of dislocations in a deformed solid, or the stick-slip motion between two mediums.

To model damage evolution from an undamaged to a damaged state, we introduce a yield stress $\sigma_y$ and corresponding yield strain $\varepsilon_y$. If the stress is less than the yield stress $\sigma \leq \sigma_y$ there is no damage and linear elasticity is applicable. If the stress is greater than the yield stress $\sigma > \sigma_y$, damage occurs and fiber failure occurs to model this irreversible behavior. We consider that a brittle solid obeys linear elasticity for stresses in the range $0 \leq \sigma \leq \sigma_y$. We also assume that Hooke's law is applicable so that the dependence of stress $\sigma$ on strain $\varepsilon$ is given by $\sigma = E_0 \varepsilon$, where $E_0$ is Young's modulus, a constant. From this equation, the corresponding yield strain is given by $\varepsilon_y = \sigma_y/E_0$.

The standard approach to the dynamic time-dependent failure of a fiber bundle is to specify an expression for the rate of failure of fibers (Coleman 1956, 1958; Newman and Phoenix 2001; Turcotte et al. 2003). The form of this breakdown rule is given by

$$\frac{dn_f(t)}{dt} = -\nu(\sigma)n_f(t), \tag{2}$$

where $n(t)$ is the number of original fibers that remain unbroken at time $t$ and $\nu(\sigma)$ is known as the hazard rate, which is a function of the fiber stress $\sigma(t)$.

It is also necessary to prescribe the dependence of the hazard rate ν on the fiber stress σ. Turcotte and Glasscoe (2004) and Nanjo and Turcotte (2015) used the equation $\nu(\sigma) = \nu_f [\sigma(t)/E_0]^\beta$, where $\nu_f$ is the reference hazard rate corresponding to a stress equal to Young's modulus of each fiber $E_0$ and β is constant. For composite materials, it was empirically found that the values of β fall in the range of 2–5 (Newman and Phoenix 2001). We modify their results to include yield stress $\sigma_y$:

$$\nu(\sigma) = \nu_f \left(\frac{\sigma - \sigma_y}{E_0}\right)^\beta \qquad \text{if } \sigma > \sigma_y. \tag{3}$$

If $0 < \sigma \leq \sigma_y$, $\nu(\sigma) = 0$ and no fiber failure occurs.

Eqs. 2 and 3 show variability in the time delay, even if the stress on the bundle is carried equally by all fibers. A fundamental question is if the cause of the time delay is associated with the damage. This cause is in a close association with the thermal fluctuations in phase changes. The temporal delay of the damage is that it takes time to nucleate micro-cracks. An in-depth discussion is given in Zapperi et al. (1997), Moreno et al. (2001), Shcherbakov and Turcotte (2003), and Kovács et al. (2008).

In order to illustrate the failure of a simple fiber-bundle under uniform loading that introduces yield stress $\sigma_y$ and yield strain $\varepsilon_y = \sigma_y/E_0$, we consider two examples without fiber replacement. The first is the instantaneous application of a uniform strain $\varepsilon_0$ to a fiber-bundle. We show that catastrophic failure in a finite time period does not occur. The second is the instantaneous application of a constant tensional force $F_0$ to a fiber-bundle. The solution obtained explains for the occurrence of a catastrophic failure of the fiber-bundle.

We first consider the case in which a uniform strain $\varepsilon_0$ is applied at time $t = 0$ and maintained upon the fiber-bundle for $t > 0$. In this case, the stress on each fiber has a constant value $\sigma_0 = E_0\varepsilon_0$. From Eq. 3 with $\sigma = \sigma_0$ (= $E_0\varepsilon_0$), the hazard rate is $v = v_f(\varepsilon_0 - \varepsilon_y)^\beta$ which is independent of time $t$ and dependent of the excess strain $\varepsilon_0 - \varepsilon_y$. Eq. 2 can be integrated to give $n_f(t) = n_0 exp\left[-v_f(\varepsilon_0 - \varepsilon_y)^\beta t\right]$, where the initial condition $n_f(0) = n_0$ has been used. The total excess force $F(t)$ carried by the fiber bundle at time $t$ is given by $F(t) = n_f(t)a(\sigma_0 - \sigma_y)$, where $a$ is the area of a fiber and $\sigma_0 - \sigma_y$ is the excess stress. The total excess force is given by $F(t) = n_0 a E_0(\varepsilon_0 - \varepsilon_y)exp\left[-v_f t(\varepsilon_0 - \varepsilon_y)^\beta\right]$. Because there is no fiber replacement, the total excess force $F(t)$ decreases as fibers fail. Catastrophic failure in a finite period of time does not occur.

Next, we consider the case in which a constant tensional force is applied to the fiber-bundle at time $t = 0$. Similar to the first case, we take no fiber replacement into account. The initial stress on each fiber at $t = 0$ is given by $\sigma_0 = F_0/n_0 a$. The applied tensional force remains constant, so that when a fiber fails the force carried by that fiber is redistributed to other fibers. Thus, the stress on surviving fibers increases with time $t$. This is uniform load sharing and is a mean-field approximation. One implication of this assumption is that all the remaining fibers have the same stress $\sigma(t)$. Equating the total excess force at time $t$ to the initial total excess force at $t = 0$, the stress on the surviving fibers is related to the number of sound fibers $n_f(t)$ by $\sigma(t) = \frac{n_0}{n_f}(\sigma_0 - \sigma_y) + \sigma_y$. Substituting this equation and Eq. 3 into Eq. 2 and integrating with the initial condition $n_f(0) = n_0$, we obtain $n_f(t) = n_0(1 - t/t_c)^{\frac{1}{\beta}}$, where $t_c$ is the time to failure of a fiber-bundle given by $t_c = v_f^{-1}\beta^{-1}(\varepsilon_0 - \varepsilon_y)^{-\beta}$. Catastrophic failure occurs at $t = t_c$. The stress in each of the remaining fibers at time $t$ is obtained by substituting $n_f(t) = n_0(1 - t/t_c)^{\frac{1}{\beta}}$ into $\sigma(t) = \frac{n_0}{n_f}(\sigma_0 - \sigma_y) + \sigma_y$ with the result $\sigma(t) = (\sigma_0 - \sigma_y)(1 - t/t_c)^{-\frac{1}{\beta}} + \sigma_y$. Each fiber satisfies linear elasticity until it fails, $\varepsilon(t) = \sigma(t)/E_0$. Substitution of $\sigma(t) = (\sigma_0 - \sigma_y)(1 - t/t_c)^{-\frac{1}{\beta}} + \sigma_y$ into $\varepsilon(t) = \sigma(t)/E_0$ gives the strain $\varepsilon(t)$ of each remaining fiber, $\varepsilon(t) = (\epsilon_0 - \varepsilon_y)(1 - t/t_c)^{-\frac{1}{\beta}} + \varepsilon_y$ where $\varepsilon_0 = \sigma_0/E_0$ and $\varepsilon_y = \sigma_y/E_0$. The stress and strain of each remaining fiber approach infinity as the time approaches the time to failure at $t = t_c$, which is finite.

2.2 Non-Newtonian Viscous Rheology Model

We consider the uniform extension of a rod that is made up of $n_0$ fibers, in which the rod is being extended at a constant strain rate $\dot{\varepsilon}$. The statistical distribution of fiber lifetimes was determined. We assume that all $n_0$ fibers have stress equal to $\sigma_y$ at $t = 0$. The stress in each fiber at subsequent times is given by $\sigma(t) = \sigma_y + E_0 t \dot{\varepsilon}$. Substitution of this equation and Eq. 3 into Eq. 2 and integration with the initial condition $n_f(0) = n_0$ give $n_f(t) = n_0 \exp[-v_f \dot{\varepsilon}^\beta t^{\beta+1}/(\beta + 1)]$. We rewrite this form with

the relation

$$n_f(\tau) = n_0 exp\left[-\frac{\tau^{\beta+1}}{\beta+1}\right], \tag{4}$$

where $\tau$ is the non-dimensional time given by $\tau = v_f^{\frac{1}{\beta+1}} \dot{\varepsilon}^{\frac{\beta}{\beta+1}} t$. $n_f(\tau)$ is shown in Fig. 1a as a function of $\tau$ while assuming several values of β. High values of β show that $n_f(\tau)$ quickly decreases with $\tau$, showing quick damage evolution. If σ ≤ σ_y, $n_f(\tau) = n_0$, then there is no damage evolution. The non-dimensional probability distribution function $f(\tau)$ for the distribution of fiber lifetime is given by

$$f(\tau) = -\frac{1}{n_0}\frac{dn_f(\tau)}{d\tau} = \tau^\beta exp\left[-\frac{\tau^{\beta+1}}{\beta+1}\right]. \tag{5}$$

This is a Weibull probability density function. A well-known case equation to show the mean non-dimensional fiber lifetime $\bar{\tau}$ is given by

$$\bar{\tau} = \int_0^\infty f(\tau)\tau d\tau = (\beta+1)^{\frac{1}{\beta+1}}\Gamma\left(\frac{\beta+2}{\beta+1}\right), \tag{6}$$

where $\Gamma(z)$ is the gamma function of $z$ (Abramowitz and Stegun 1965). We now use the form of $f(\tau)$ and the values of $\bar{\tau}$ for various values of β. The relation between $f(\tau)$ and τ (Eq. 5) is given in Fig. 1b for several values of β. This figure illustrates that the spread of the function $f(\tau)$ increases as the value of β decreases. Fig. 1c represents Eq. 6 relating $\bar{\tau}$ to β. The lifetime $\bar{\tau}$ decreases as β increases.

The dimensional mean fiber lifetime $\bar{t}$ is obtained by substation of Eq. 6 into $\tau = v_f^{\frac{1}{\beta+1}} \dot{\varepsilon}^{\frac{\beta}{\beta+1}} t$ to give

$$\bar{t} = \left(\frac{\beta+1}{v_f}\right)^{\frac{1}{\beta+1}} \Gamma\left(\frac{\beta+2}{\beta+1}\right) \dot{\varepsilon}^{-\frac{\beta}{\beta+1}}. \tag{7}$$

$\bar{t}$ is shown in Fig. 1d as a function of $\dot{\varepsilon}$, assuming $v_f = 1$ and several values of β. For large β values, $\bar{t}$ is proportional to the inverse of $\dot{\varepsilon}$.

In order to get the mean stress on the fiber bundle, we consider the mean stress on a single fiber that has been replaced many times (Fig. 2). For this single fiber, we assume that our replacement hypothesis for the fiber bundle described above is applicable. When this fiber fails, it is instantaneously replaced by a new fiber with stress equal to the yield stress $\sigma_y$. Time intervals between failure events satisfy the probability distribution of the lifetime given by Eq. 5. The strength of a newly created fiber is proportional to the period from the time at which it is replaced to the time at which it failed. The fiber strength before and after replacement is not related. The fiber stress is $\sigma = \sigma_y$ at $t = 0$, and the mean stress during the period from 0 to $t$ on a fiber $\tilde{\sigma}$ is given by $\tilde{\sigma} = E_0 t\dot{\varepsilon}/2 + \sigma_y$.

The mean stress on the fiber after many replacements $\bar{\sigma}$ is given by

$$\bar{\sigma} = \int_0^\infty (\tilde{\sigma} - \sigma_y)\frac{t}{\bar{t}}f(t)dt + \sigma_y = \frac{E_0\dot{\varepsilon}}{2\bar{t}}\int_0^\infty t^2 f(t)dt + \sigma_y. \tag{8}$$

Using $\tau = v_f^{\frac{1}{\beta+1}}\dot{\varepsilon}^{\frac{\beta}{\beta+1}}t$, we derive

$$\frac{2(\bar{\sigma}-\sigma_y)}{E_0} = \left(\frac{\dot{\varepsilon}}{v_f}\right)^{\frac{1}{\beta+1}}\frac{1}{\bar{\tau}}\int_0^\infty \tau^{\beta+2}\exp\left(-\frac{\tau^{\beta+1}}{\beta+1}\right)d\tau.$$

$$= \left(\frac{\beta+1}{v_f}\dot{\varepsilon}\right)^{\frac{1}{\beta+1}}\Gamma\left(\frac{\beta+3}{\beta+1}\right)/\Gamma\left(\frac{\beta+2}{\beta+1}\right). \tag{9}$$

This is rewritten as

$$\dot{\varepsilon} = \frac{1}{\tau_c}\left(\frac{\bar{\sigma}-\sigma_y}{E_0}\right)^n, \tag{10}$$

where $n = \beta + 1$ and $\tau_c$ is a characteristic time given by $\tau_c = \left(\frac{\beta+1}{v_f}\right)\left[\frac{1}{2}\Gamma\left(\frac{\beta+3}{\beta+1}\right)/\Gamma\left(\frac{\beta+2}{\beta+1}\right)\right]^{\beta+1}$.

Each fiber in the bundle behaves like the single fiber we have considered. We further assume that the temporal distribution of the mean stress for a large number of fibers is equal to that for the single fiber considered. Thus Eq. 10 is regarded as the expression relating the mean stress $\bar{\sigma}$ on our fiber bundle to the applied strain rate $\dot{\varepsilon}$. This is a non-Newtonian viscous rheology. This form of damage rheology is equivalent to the form of rheology (Eq. 1) that is derived from considerations of dislocation densities and atomic diffusivities. An advantage of the former rheology in Eq. (10) over the latter one (Eq. 1) is to provide justification for the use of non-Newtonian viscous flow for the continuum deformation of brittle materials. The rheology in Eq. (10) has applicability to cases where materials are too cold to justify the use of the rheology in Eq. 1 associated with dislocation creep and diffusion creep, both of which are thermally activated.

The characteristic time $\tau_c$ is now related to the hazard rate $v_f$. Physically, this is the reference of the delay time associated with damage evolution. However, Nanjo and Turcotte (2005) empirically fitted it to observed rheology in the Earth's crust.

**3 Application to viscoelasticity**

It is useful for a number of problems to combine a fluid rheology on a long timescale with elastic behavior on a short timescale. For this purpose, viscoelastic rheology is usually used. The models, which include the Maxwell model, the Kelvin-Voigt model and their generalized versions, are used to predict a material's response under different loading conditions. As an application example of our rheological model in Eq. 10, we consider the Maxwell viscoelasticity. We further consider a case where a constant strain is applied to the viscoelastic medium. The response of the medium is applicable to an understanding of the time-dependent decay of aftershocks. Detailed discussion of the application to viscoelasticity is given in Nanjo et al. (2005).

The Maxwell model for viscoelasticity considers a material in which the total strain rate $\dot{\varepsilon}_{ve}$ is hypothesized to be the sum of an elastic strain rate $\dot{\varepsilon}_{el}$ and viscous strain rate $\dot{\varepsilon}_v$ given as $\dot{\varepsilon}_{ve} = \dot{\varepsilon}_{el} + \dot{\varepsilon}_v$. The elastic strain of the material is $\varepsilon_{el} = \bar{\sigma}/E_0$ and the time derivative is $\dot{\varepsilon}_{el} = \frac{1}{E_0}\frac{d\bar{\sigma}}{dt}$. If the stress on the medium is less than the yield stress $\bar{\sigma} < \sigma_y$ there is no viscoelastic deformation and the material behaves with elasticity. If the stress on the medium is greater than the yield stress $\bar{\sigma} > \sigma_y$ viscous strain will occur and the viscous strain rate is given by Eq. 10 as $\dot{\varepsilon}_v = \frac{1}{\tau_c}\left(\frac{\bar{\sigma}-\sigma_y}{E_0}\right)^n$. We then have

$$\dot{\varepsilon}_{ve} = \frac{1}{E_0}\frac{d\bar{\sigma}}{dt} + \frac{1}{\tau_c}\left(\frac{\bar{\sigma}-\sigma_y}{E_0}\right)^n. \qquad (11)$$

This is the rheological law relating strain rate, stress, and the rate of change of stress for our Maxwell viscoelastic material.

We consider the viscoelastic medium to which a constant strain has been applied. A strain $\varepsilon_0 > \varepsilon_y$ is applied instantaneously at $t = 0$ and is held constant. The behavior of the material is elastic during very rapid application of the strain. Thus, the initial stress $\sigma_0$ is $\sigma_0 = E_0\varepsilon_0$. If $\varepsilon_0 \leq \varepsilon_y$, no damage occurs and the initial stress remains unchanged $\bar{\sigma} = \sigma_0$ for $t > 0$. If $\varepsilon_0 > \varepsilon_y$, the material is strained elastically along the path AB in Fig. 3. The total strain is then maintained constant $\varepsilon_0$ so that damage evolves and repetitive failures occur. Due to the damage and failures, the stress on the sample relaxes from the initial stress $\sigma_0$ to the yield stress $\sigma_y$. This relaxation takes places along the path BCD illustrated in Fig. 3. This solution will give the time dependence of stress $\sigma(t)$ during stress relaxation.

For $t > 0$, $\dot{\varepsilon}_{ve} = 0$ and Eq. 11 reduces to

$$0 = \frac{1}{E_0}\frac{d\bar{\sigma}}{dt} + \frac{1}{\tau_c}\left(\frac{\bar{\sigma}-\sigma_y}{E_0}\right)^n. \tag{12}$$

Integration with the initial condition $\bar{\sigma} = \sigma_0$ at $t = 0$ gives

$$\frac{\bar{\sigma}-\sigma_y}{\sigma_0-\sigma_y} = \frac{1}{\left\{1+(n-1)\left(\frac{\sigma_0-\sigma_y}{E_0}\right)^{n-1}\left(\frac{t}{\tau_c}\right)\right\}^{1/(n-1)}}. \tag{13}$$

In the limit $t \to \infty$, the result is $\bar{\sigma}(\infty) = \sigma_y$. The normalized stress $(\bar{\sigma} - \sigma_y)/(\sigma_0 - \sigma_y)$ is given as a function of non-dimensional time $t/\tau_c$ in Fig. 4a taking into account the normalized excess stress $(\sigma_0 - \sigma_y)/E_0 = 1.0$ and several values of $n$. Increasing the values of $n$ greatly slows stress relaxation. The normalized stress $(\bar{\sigma} - \sigma_y)/(\sigma_0 - \sigma_y)$ is given as a function of $t/\tau_c$ in Fig. 4b taking the exponent $n = 1$ and several values of $(\sigma_0 - \sigma_y)/E_0$. High initial stresses relax quickly followed by a power-law relaxation.

This relaxation process is applicable to an understanding of the time-dependent decay of aftershocks that follow the main shock. A universal scaling law applicable to the temporal decay of aftershock activity is known as the Omori-Utsu law (Omori 1894; Utsu 1961). The most widely used form is given as

$$\frac{dN}{dt} = \frac{K}{(c+t)^p}. \qquad (14)$$

where $N$ is the number of aftershocks with magnitude greater than a specified value, $c$ and $K$ are constants, and the power-law exponent $p$ has a value somewhat greater than unity.

Following Shchervakov and Turcotte (2003) and Shcherbakov et al. (2005), our working hypothesis is that stress transfer during a main shock increases the stress $\bar{\sigma}$ and strain $\varepsilon_{ve}$ above the yield values $\sigma_y$ and $\varepsilon_y$ in some regions adjacent to the fault on which the main shock occurred. The increases to stress and strain are essentially instantaneous and follow linear elasticity. We believe that it is a good approximation to neglect any increase in regional stress due to tectonics during the aftershock sequence. We also neglect any increase in stress due to large aftershocks because only 3% of the total energy is associated with the aftershock sequence while 97% is associated with the main shock (Nanjo and Nagahama 2000; Shcherbakov et al. 2005). We hypothesize that the applied strain $\varepsilon_0$ remains constant and that aftershocks relax the stress $\bar{\sigma}$ to its yield value $\sigma_y$. The occurrence of aftershocks is attributed to this relaxation process as given in Eq. 13. The time delay of aftershocks relative to the main shock is directly analogous to the time delay of the damage. This is because it takes time to nucleate micro-cracks, i.e. aftershocks.

In order to quantify the rate of aftershock occurrence, we determined the rate of energy release in the relaxation process. The stored elastic energy release (per unit mass) $e_0$ in a material after an instantaneous strain $\varepsilon_0$ has been applied along path AB is $e_0 = E_0 \varepsilon_0^2 / 2$. Since the strain is constant during stress relaxation, no work is done on the sample. If the applied strain (stress) is instantaneously removed at point C, we hypothesize that the sample will follow the elastic path CF that is parallel to path AB. The elastic energy $e_1$ recovered during stress relaxation on this path is given by $e_1 = \bar{\sigma} \varepsilon_0 / 2$. We assume that the difference between the energy added $e_0$ and the energy recovered $e_1$ is lost in aftershocks. This energy $e_{AF}$, which is given by $e_{AF} = e_0 - e_1$, corresponds to the area ABCF. The total energy of aftershocks $e_{AFT}$ obtained in the limit $t \to \infty$ corresponds to the area ABCDEF with the result $e_{AFT} = E_0 \varepsilon_0 (\varepsilon_0 - \varepsilon_y)/2$. Using this equation and integrating the time derivative of $e_{AF} = e_0 - e_1$, we obtain the rate of energy release

$$\frac{1}{e_{AFT}} \frac{de_{AF}}{dt} = \frac{\left(\frac{1}{n-1}\right) c^{\frac{1}{n-1}}}{(c+t)^{\frac{n}{n-1}}} = \frac{(p-1)c^{p-1}}{(c+t)^p}. \tag{15}$$

where we take $n = \frac{p}{p-1}$ and $c = \frac{\tau_c}{(n-1)(\varepsilon_0-\varepsilon_y)^{n-1}}$. See Appendix 1 for detailed derivation of Eq. 15.

Following Newman et al. (1995), Shcherbakov and Turcotte (2003), Turcotte and Glasscoe (2004), and Shchebakov et al. (2005), we hypothesize that the rate of energy release is equal to the rate of occurrence of earthquakes $de_{AF}/dt = dN/dt$. If we assume $K = e_{AFT}(p-1)c^{p-1}$, Eq. 15 is identical to Eq. 14.

There are several attempts to explain aftershock decay patterns within damage models (e.g. Ben-Zion and Lyakhovsky, 2006) and within the context of rate-and-state dependent frictional rheology (e.g. Dieterich 1994; Kaneko and Lapusta 2008). What new feature that our model introduces into this problem is to associate aftershock relaxation, i.e., months to years with the long-term crustal deformation, i.e., millions of years. The behavior of the deforming the Earth's crust can be modeled as the non-Newtonian viscous flow in Eq. 10. Once a large earthquake (a main shock) occurs in the crust, stress suddenly increases in some regions adjacent to the fault on which the main shock occurred. Stress relaxation is accompanied by the aftershock sequence. Using a viscoelastic version of our model, we got the Omori-Utsu law temporal aftershock decay in Eq. 15. The "healing" (modeled by fiber replacement in our study), needed for the continuum deformation of the crust, is introduced into the problem of aftershock decay.

**4 Discussion**

The basis of our mechanism is the application of a renewable FBM to brittle deformations. FBM has been applied successfully to the failure of composite materials. This model, as defined in Eqs. 2 and 3, is inherently dependent on time through the hazard rate. This time dependence is associated with the nucleation and coalescence of micro-cracks. In order to represent the continuum deformation of a solid, it is necessary to introduce "healing". Following Zapperi et al. (1997), Kun et al. (2000), Moreno et al. (2000), Kun et al. (2006), and Halász and Kun (2009), we do this by replacing a failed fiber by a new fiber with stress equal to yield stress. The result is a non-Newtonian viscous rheology as previously found by Turcotte and Glasscoe (2004) and Nanjo and Turcotte (2005). However, fiber replacement is not damage mechanics but depends on how to set an ad-hoc boundary condition to constrain the solution of the model.

Continuum damage in the context of fiber bundle with multiple failure events allowed have been addressed including stick slip generation by Kun et al. (2006) (see also Halász and Kun 2009). The feature introduced by their research is the introduction of the sudden stiffness degradation. These authors assumed that at the failure point of a fiber, the stiffness of the fiber gets reduced. The obtained relation between the average load on a fiber and the strain shows plasticity response during the period of multiple failure events allowed. Despite mechanistic details, the microscopic processes and the macroscopic constitutive behavior look very similar between our research and their research. Our assumption that single fibers have linear elastic behavior up to fiber failure, which is similar to Kun et al. (2006). We derived and used the Weibull probability density function for fiber lifetimes in Eq. 5, which is equivalent to the equation assumed for the disordered breaking thresholds in Kun et al. (2006). We calculated the mean stress on the fiber after many replacements $\bar{\sigma}$ in Eq. 8, which is similar to the average load on a fiber in Kun et al. (2006). They show, by analytical calculation, that plastic response of their fiber bundle model on the macro-scale emerges, which is the same as Eq. 10 with assuming yield stress $\sigma_y = 0$ in the limit of the power-law exponent *n* to be infinity. The similarity between Kun et al. (2006) and our study suggests that incorporating physical evidence of healing, such as the sudden stiffness degradation, into our model is an important theoretical development that is worthy of further exploration.

Following Shcherbakov and Turcotte (2003), Shcherbakov et al. (2005), Nanjo et al. (2005), and Manaker et al. (2006), we introduced the concept of yield stress into FBM. As a consequence, the power-law rheology (Eq. 10) relating between strain rate $\dot{\varepsilon}$ and excess stress $\bar{\sigma} - \sigma_y$ is obtained. This theoretical development expands upon our previous works (Turcotte and Glasscoe 2004; Nanjo and Turcotte, 2005) that did not consider yield stress. This is different from normal constitutive creep laws such as in Eq. 1 without yield stress. However, an empirical equation similar to Eq. 10 has been used in engineering to predict more accurately the rheology of deformation with stress above yield stress. This is called the yield-power law (YPL) model (Hamphill et al. 1993; Houwen and Greehan 1986; Skelland 1967; Reed and Phiehvari 1993; Zamora and Lord 1974) that has been applied to predict the rheological behavior of ductile materials such as mud. Comparison with the YPL models shows that the power-law rheology (Eq. 10) relating between strain rate $\dot{\varepsilon}$ and excess stress $\bar{\sigma} - \sigma_y$ is appropriate to precisely predict both the continuum deformation of ductile materials and the continuum deformation of brittle materials.

We now compare the results derived above with a laboratory experiment and geophysical observations. Yamaguchi and others (Yamaguchi et al. 2009; Morishita et al. 2010) conducted an experiment of continuum shear deformation of a material that consists of two elastic bodies. Two elastic bodies, when slid against each other, exhibited stick-slip motion repeating lock and sliding. A constant pull velocity $V_{\text{pull}}$ was applied to one body while the other remained fixed. Tensile force (frictional force) $F$ over many occurrences of stick-slips was monitored for several constant pull velocities from $V_{\text{pull}} = 1$ to $V_{\text{pull}} = 1000$ μm/s. No clear stick-slip behavior was observed below $V_{\text{pull}} = 200$ μm/s. Above this value, stick-slip behavior was observed: $F$ increased with time $t$ (lock) followed by a sudden force drop (sliding). The average force over many stick-slip motions $\bar{F}$ was correlated to the velocity $V_{\text{pull}}$, given by $V_{pull} \propto \bar{F}^{1/0.15}$. This result is in good agreement with Eq. 10 taking $n = 6.7$ if $\bar{\sigma}/\sigma_y \gg 1$.

Houseman and England (1986) and England and Molnar (1997) considered an indenter model for continental deformation, which significantly takes place on faults (e.g. King 1983; King et al. 1994; Thatcher 1995; Jackson 2002). England and others applied the finite element method to a thin non-Newtonian viscous sheet to obtain solutions. Using the power-law rheology given in an equation similar to Eq. 10, they obtained results for $n = 3$ and 10. These authors compared their results with observations in the Indian-Asian collision zone and found broad agreement provided that the power-law exponent is large ($n > 3$). Thus, the use of a non-Newtonian viscous flow rheology given as in Eq 10 is common to modelling the continuum deformation of brittle materials in a wide range of scales: from a laboratory experiment with stick-slip motion to orogenies such as the Indian-Asian collision with faulting.

The normal constitutive law in Eq. 1 is also derived from considerations of atomic diffusivities and dislocation densities. The exponent is $n = 1$ for diffusion creep and $n = 3$-5 for dislocation creep. The large value of the exponent for the laboratory experiment ($n = 6.7$) and broad agreement of the large values of the exponent for the continental deformation ($n > 3$) are certainly not surprising for aftershocks. Nanjo et al. (2007) studied the decay of aftershock activity for four Japanese earthquakes: 1995 in Kobe (magnitude $M = 7.3$), 2000 in Tottori ($M = 7.3$), 2004 in Niigata ($M = 6.8$), 2005 in Fukuoka ($M = 7.0$). These authors determined the rates of occurrence of aftershocks in numbers per day as a function of time for 1000 days (378 days for Niigata and 225 days for Fukuoka) after a main shock. They fitted an equation similar to Eq. 14 to the data and found $p = 1.19$~$1.23$ for Kobe, $p = 1.21$~$1.24$ for Tottori, $p = 1.32$~$1.34$ for Niigata, and $p = 1.28$~$1.29$ for Fukuoka. Shcherbakov et al. (2005) studied aftershock sequences following four California earthquakes and found $p = 1.22 \pm 0.03$ for 1992 in Landers ($M = 7.3$), $p = 1.18 \pm 0.02$ for 1994 in Northridge ($M = 6.7$), $p = 1.21 \pm 0.05$ for 1999 in Hector Mine ($M = 7.1$), and $p = 1.12 \pm 0.02$ for 2003 in San Simeon ($M = 6.5$). From $n = p/(p - 1)$, we found that the values of the exponent $n$ fell in the range $n = 4 \sim 11$. Reasenberg and Jones (1989) also carried out a detailed study of aftershocks for major earthquakes in California and found the mean value $p = 1.07 \pm 0.03$. From $n = p/(p - 1)$, we found $n = 15$. Moreno et al. (2001) used a large value ($\beta = 30$) to perform their simulations of a fiber-bundle model for complex aftershock sequences. From $n = \beta + 1$, we observed $n = 31$. Although applicable values of $n$ are not well constrained, one important aspect of our study is that the power-law exponent $n$ is likely large for the continuum deformation of brittle solids.

In the limit of the power-law exponent $n$ to infinity, there is no dependence of stress on strain rate, and the rheology is perfectly plastic. The large power-law exponents we found show the behavior of the deforming materials approaches that of a perfectly plastic material. From $\beta = n - 1$ and Eq. 5 (Fig. 1a) showing that the spread of fiber-lifetime distribution decreases with increasing $\beta$, the physics behind this perfect plasticity dictates that recurrence of failures approaches perfectly periodic. This is applicable to a frictional rheology without taking frictional hysteresis into consideration, such as the Anderson theory of faulting (Anderson 1951; Scholz 2002). The continuum deformation of brittle solids modeled by a power-law rheology in Eq. 10 with large $n$ values shows intermediate between perfectly plastic rheology based on periodic recurrence failures and power-law rheology with normal range of exponents (typically, $n = 2\text{-}5$: Newman and Phoenix 2001) based on randomized recurrence failures.

Some forms of "damage" that we did not consider in this paper are clearly thermally activated. The deformation of solids by diffusion and dislocation creep is an example. The ability of vacancies and dislocations to move through a crystal is governed by an exponential dependence on absolute temperature. Another example is given by Nakatani (2001), who documented a systematic temperature dependence of rate and state friction. Sornette and Ouillon (2005) used a thermally activated rupture process to find that seismic decay rates after main shocks following the Omori-Utsu law in Eq. 14. The continuum deformation of the continental crust has already been considered as a thermally activated process by Nanjo and Turcotte (2005) who utilized a fiber-bundle model. These authors assumed that fiber failure is a thermally activated earthquake and modified the hazard rate in Eq. 3 to a form that links the dependence of the hazard rate to the absolute temperature. They obtained a power-law relation between $\bar{\sigma}$ and $\dot{\varepsilon}$ and the rheology was exponentially dependent on the inverse absolute temperature as given by Eq. 1. Their analyses, based on laboratory experiments (e.g. Nakatani 2001), argued in favor of thermally activated damage in order to find the strength envelope of the continental lithosphere.

However, it is a matter of controversy whether temperature plays a significant role in the damage of brittle failure of materials. Guarino et al. (1998) varied the temperature in their experiments on the fracture of chipboard and found no effect.

The analysis given in this paper is for a uniaxial problem. This is the reason why the solution to this relatively simple problem is illustrative and the methodology can be readily adapted to understand the irreversible deformation of brittle materials not only in geophysics but also in engineering and material science. However, it is clearly desirable to extend the analysis to a fully three-dimensional formulation to treat relatively realistic but complicated problems. One way to do this is to reduce the shear modulus by damage but maintain the bulk modulus as an invariant. This approach has been discussed by Lyakhovsky et al. (1997) and Manaker et al. (2006).

## 5 Conclusion

This paper uses a particular version of the fiber bundle model (FBM), often used in damage mechanics, to explain the success of non-Newtonian viscous flow models for the continuum deformation of brittle materials. The particular version of the model uses a time-dependent variation on a model introduced by Coleman (1956, 1958). In particular, it introduces a yield stress, which not only provides a minimum stress level for the breakdown process to occur (a power-law in actual stress minus yield stress), but also becomes the stress carried by the fiber (rather than zero) immediately after the fiber fails. In fact, a second assumption is that the broken fiber is "replaced" with a new fiber that is already at the yield stress and which then has the same failure rate in terms of increasing stress. After considering two simple examples, the author shows that when the FBM is applied to brittle deformation of a solid, a non-Newtonian, power law, viscous rheology is obtained. The model is also used to explain stress relaxation in viscoelasticity through a stress relaxation problem. In this context the focus is on making connection to the Omori-Utsu law for the rate of aftershocks following an earthquake.

**Acknowledgements**

The author thanks the Editor K. Ravi-Chandar and two anonymous reviewers for constructive comments. A part of this study was conducted under the auspices of the MEXT Program for Promoting the Reform of National Universities.

**Appendix 1** Derivation of the equation of the rate of energy release in Eq. 15

Using the stored elastic energy $e_0 = E_0 \varepsilon_0^2 / 2$ and the recovered elastic energy $e_1 = \bar{\sigma} \varepsilon_0 / 2$, we obtain the energy lost in aftershocks $e_{AF} = \frac{E_0 \varepsilon_0^2}{2} - \frac{\bar{\sigma} \varepsilon_0}{2}$. Substitution of Eq. 13 into this equation gives

$$e_{AF} = \frac{E_0 \varepsilon_0}{2} (\varepsilon_0 - \varepsilon_y) \left[ \frac{1}{\left\{ 1 + (n-1)(\varepsilon_0 - \varepsilon_y)^{n-1} \left( \frac{t}{\tau_c} \right) \right\}^{1/(n-1)}} \right]. \tag{A1}$$

Using the total energy of aftershocks $e_{AFT} = E_0\varepsilon_0(\varepsilon_0 - \varepsilon_y)/2$, we rewrite Eq. A1 as

$$e_{AF} = e_{AFT}\left[\frac{1}{\{1+(n-1)(\varepsilon_0-\varepsilon_y)^{n-1}\left(\frac{t}{\tau_c}\right)\}^{1/(n-1)}}\right]. \tag{A2}$$

Taking the time derivative of Eq. A2, we obtain the rate of energy release

$$\frac{1}{e_{AFT}}\frac{de_{AF}}{dt} = \frac{\frac{1}{\tau_c}(\varepsilon_0-\varepsilon_y)^{n-1}}{\{1+(n-1)(\varepsilon_0-\varepsilon_y)^{n-1}\left(\frac{t}{\tau_c}\right)\}^{\frac{n}{n-1}}}. \tag{A3}$$

Using $c = \frac{\tau_c}{(n-1)(\varepsilon_0-\varepsilon_y)^{n-1}}$, we rewrite Eq. A3 as

$$\frac{1}{e_{AFT}}\frac{de_{AF}}{dt} = \frac{\left(\frac{1}{n-1}\right)c^{\frac{1}{n-1}}}{(c+t)^{\frac{n}{n-1}}}. \tag{A4}$$

If we take $n = \frac{p}{p-1}$, Eq. A4 is rewritten as

$$\frac{1}{e_{AFT}}\frac{de_{AF}}{dt} = \frac{(p-1)c^{p-1}}{(c+t)^p}. \tag{A5}$$

This is the equation of the rate of energy release in aftershocks in Eq. 15.

**Figure Captions**

**Fig. 1** (a) Dependence of the normalized number of unbroken fibers $n_f(\tau)/n_0$ on the non-dimensional time $\tau$ for several values of β. Also included for comparison is $n_f(\tau)/n_0 = 1$ for the case in which no fiber failure occurs. (b) Dependence of the non-dimensional probability density function $f(\tau)$ on the non-dimensional time $\tau$ for several values of β (modified from Nanjo and Turcotte 2005). (c) Dependence of the mean non-dimensional fiber lifetime $\bar{\tau}$ on β (modified from Nanjo and Turcotte 2005). (d) Dependence of the dimensional fiber lifetime $\bar{t}$ on the strain rate $\dot{\varepsilon}$ for several values of β taking $v_f$ = 1.

**Fig. 2** Schematic time-stress diagram associated with fiber failure events. The normalized stress $\sigma/\sigma_y$ is given as a function of time *t* for a fiber. The fiber is extended at a constant rate $\dot{\varepsilon}$. This rate linearly increases stress with time *t*. A failed fiber is replaced by a new one with stress σ_y (dashed line). The replacement of new fibers and their rupture have been randomized for sufficient times. The mean stress on the considered fiber after many replacements is given by $\bar{\sigma}$ (dash-dotted line).

**Fig. 3** Schematic stress-strain diagram for Maxwell viscoelastic material (modified from Nanjo et al. 2005). Stress $\sigma_0 > \sigma_y$ and strain $\varepsilon_0 > \varepsilon_y$ are instantaneously applied to a material at time *t* = 0. The material elastically behaves and follows the linear path AB. Subsequently the strain $\varepsilon_{ve} = \varepsilon_0$ is maintained constant and the material damage and failure associated with the non-Newtonian flow relax the applied stress $\sigma_0$ to the yield stress $\sigma_y$ along the path BCD. In order to determine the energy associated with the relaxation process, we instantaneously remove the stress $\bar{\sigma}$ and strain $\varepsilon_0$ at point C. The subsequent linear elastic behavior of the material follows the path CF. The total lost energy $e_{AFT}$ in the limit $t \to \infty$ corresponds to the area ABCDEF. The lost energy in aftershocks $e_{AF}$ corresponds to the area ABCF.

**Fig. 4** Dependence of the normalized stress $(\bar{\sigma} - \sigma_y)/(\sigma_0 - \sigma_y)$ on the normalized time $t/\tau_c$ (a) for several values of *n* with normalized initial stress $(\sigma_0 - \sigma_y)/E_0 = 1$, and (b) for several values of $(\sigma_0 - \sigma_y)/E_0$ with *n* = 3.

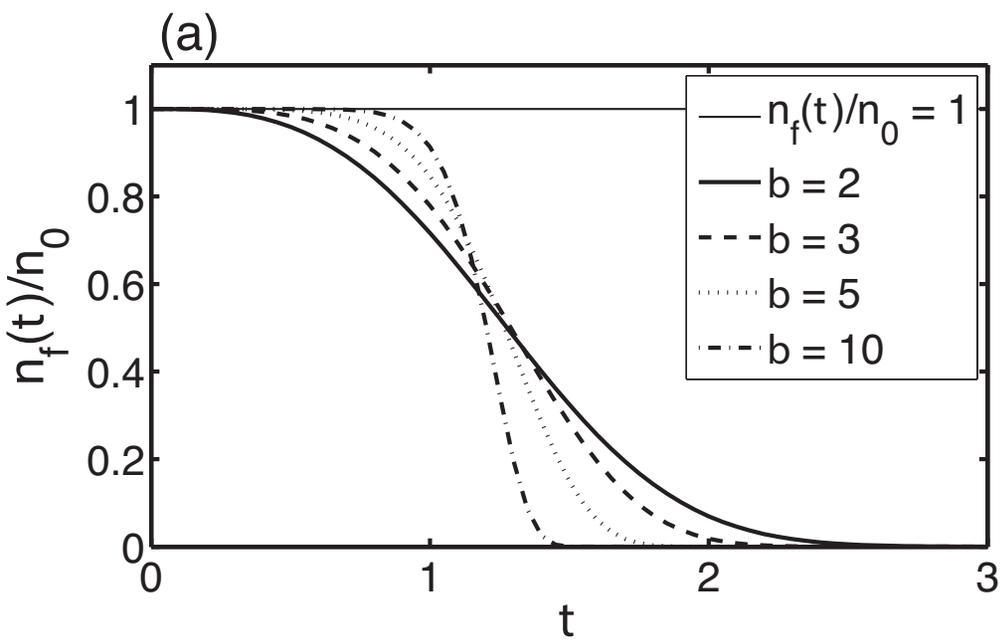
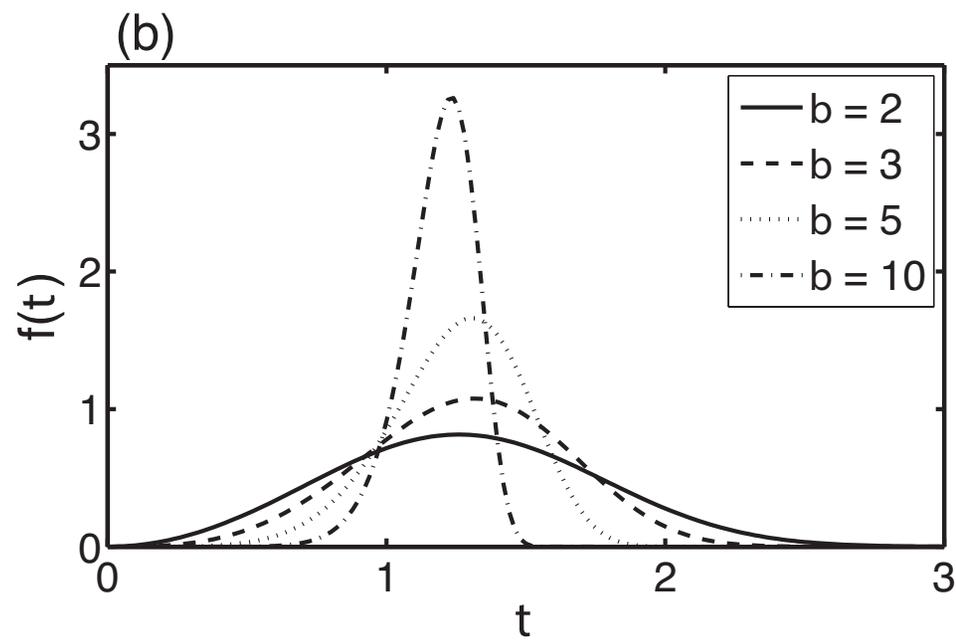
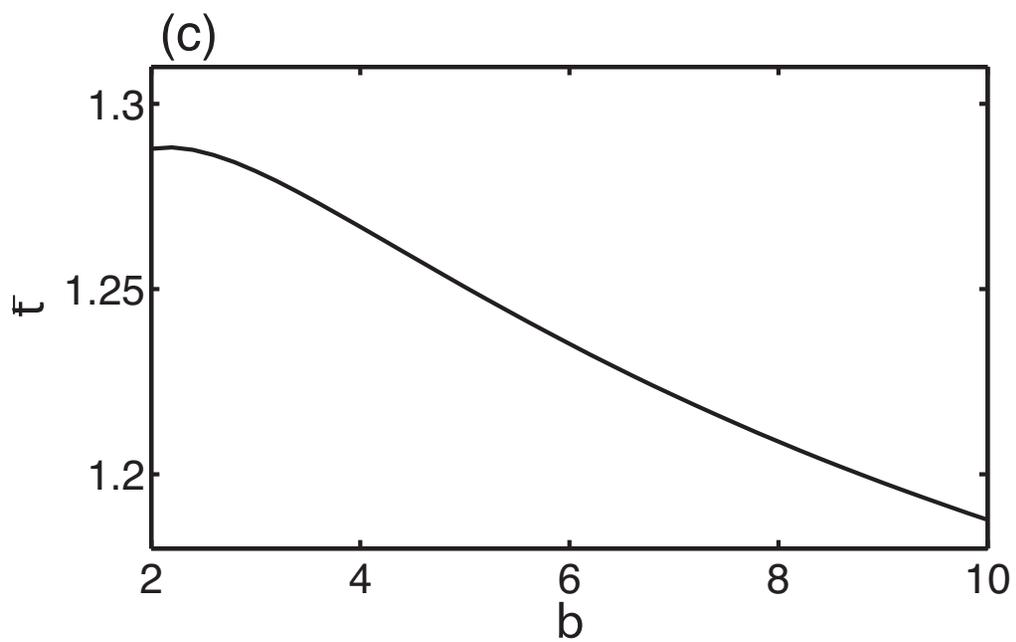
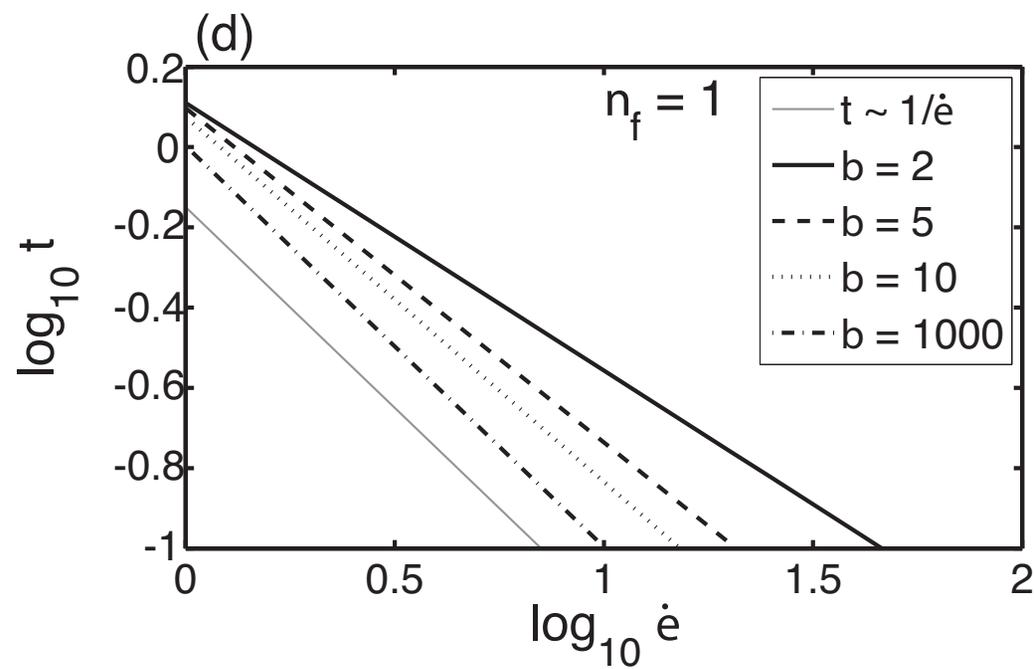

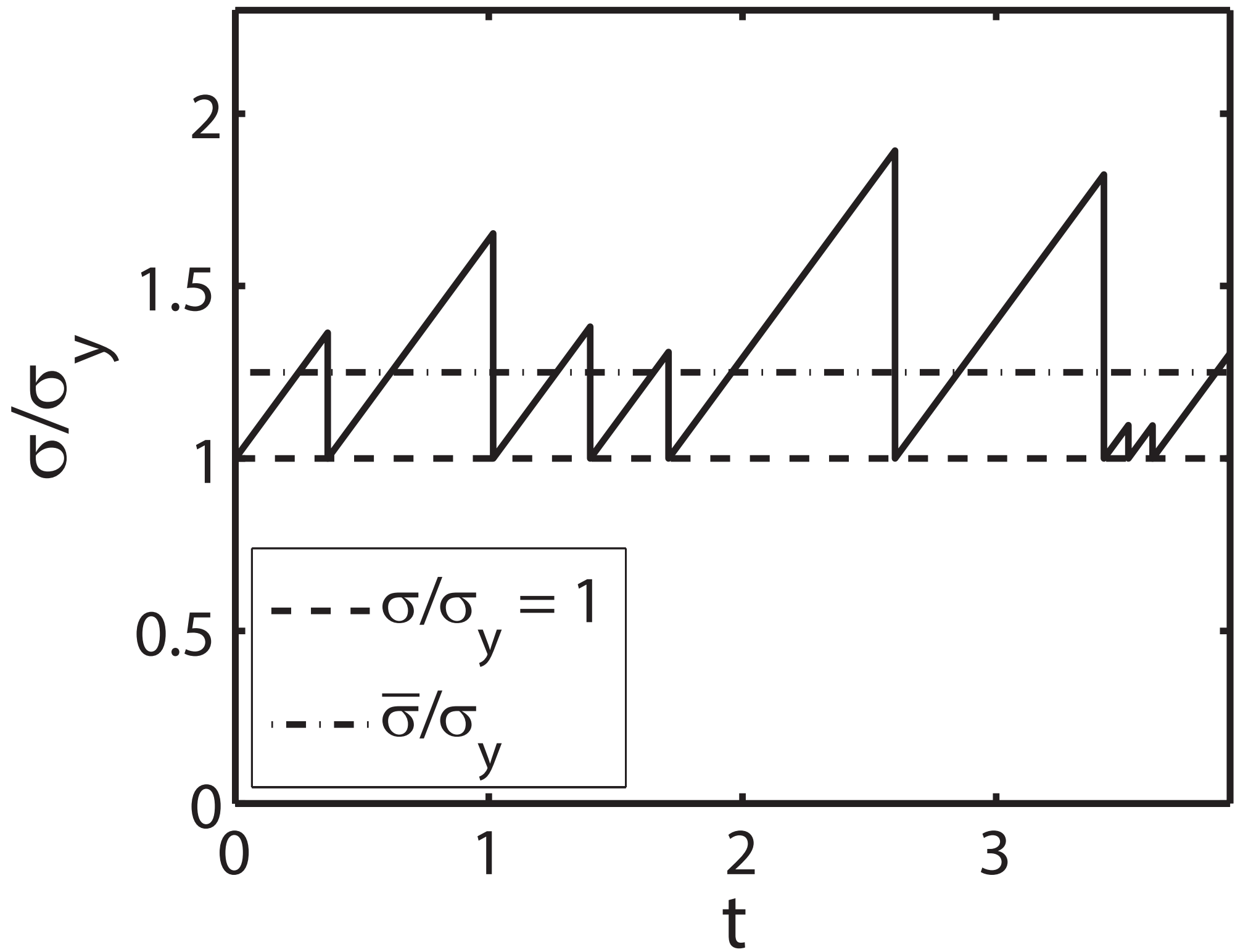

Fig2

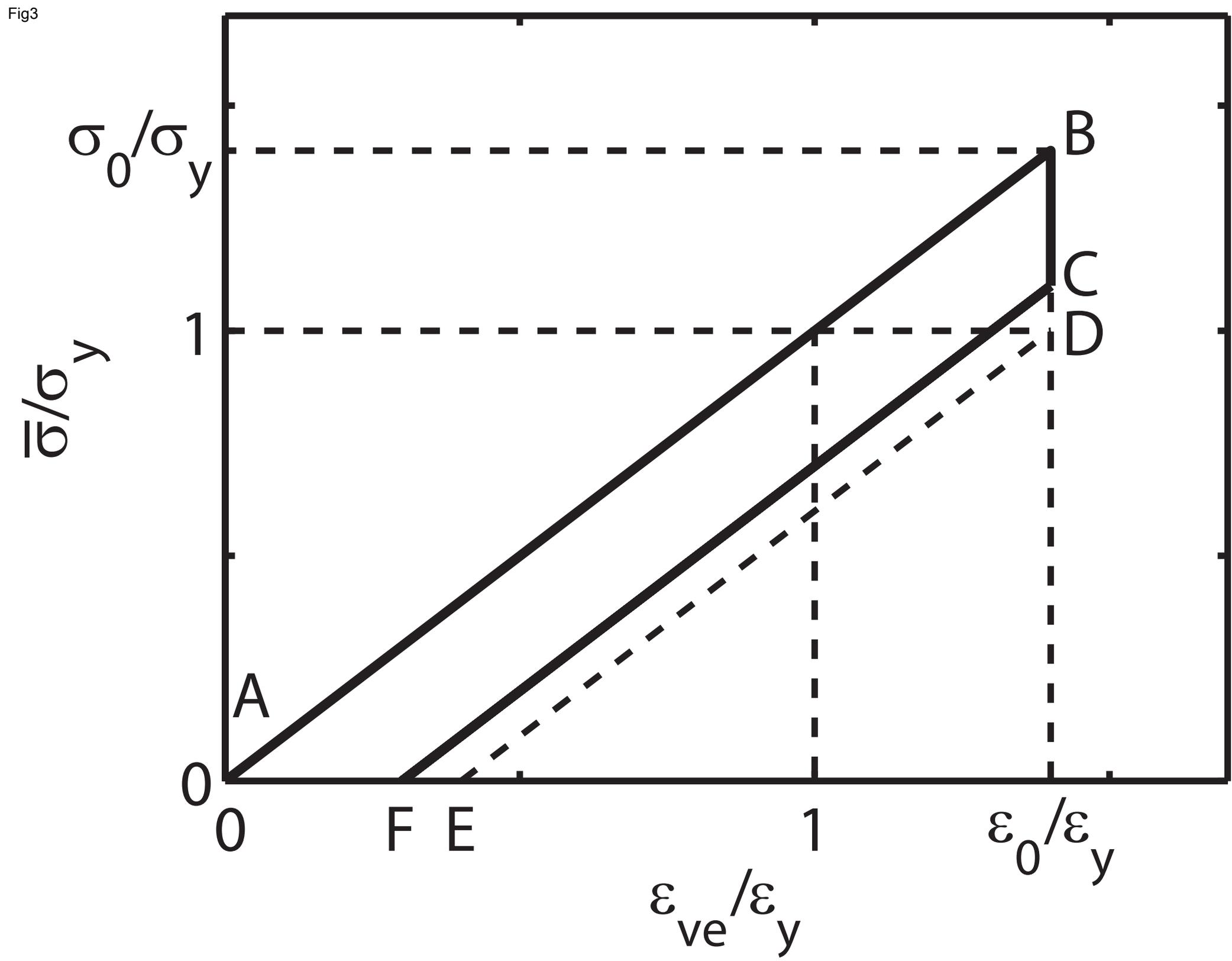

Fig3

Fig4

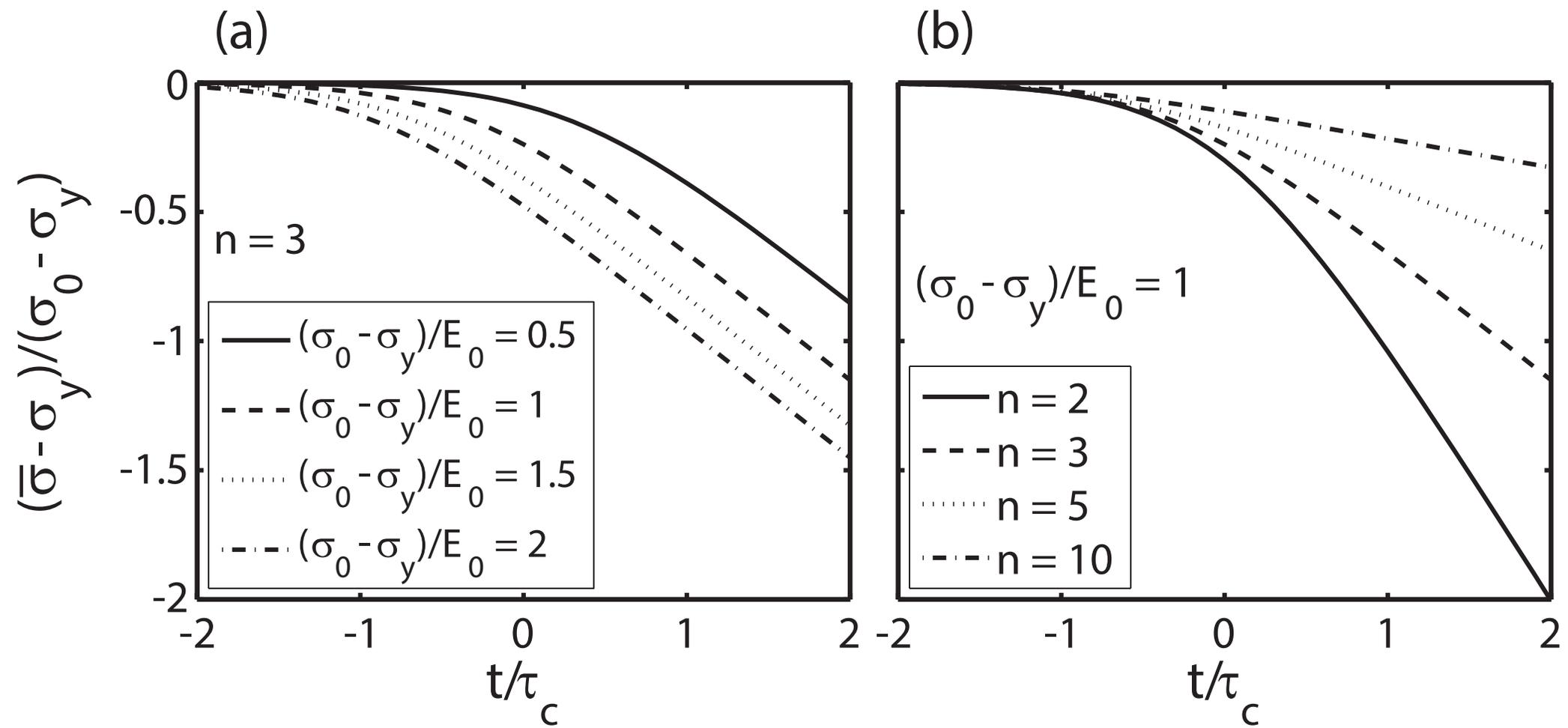